\documentclass[aps,prl,amsmath,amssymb,floatfix,twocolumn]{revtex4}
\usepackage{graphicx}
\usepackage{subfigure}
\usepackage{color}
\usepackage{hyperref}

\begin{document}

\title{Effects of Berry phase and instantons in one dimensional Kondo-Heisenberg model
}

\author{Pallab Goswami}
\affiliation{Department of Physics and Astronomy, Rice University, Houston, TX 77005}
\author{Qimiao Si}
\affiliation{Department of Physics and Astronomy, Rice University, Houston, TX 77005}

\date{\today}

\begin{abstract}
Motivated by
the global phase diagram of antiferromagnetic heavy fermion metals, we study the Kondo effect from the perspective of a nonlinear sigma model
in the  one dimensional Kondo-Heisenberg model
away from half-filling. We focus on the  effects of the instanton configurations of the sigma-model field  and the associated Berry phase.
Guided by the results derived using bosonization methods, we demonstrate that the Kondo singlet formation is accompanied by an emergent Berry phase.
This Berry phase also captures the competition between the Kondo singlet formation and spin Peierls correlations. Related effects
are likely to be realized in Kondo lattice systems in higher dimensions.

\end{abstract}


\maketitle

{\it Introduction:} Antiferromagnetic (AFM) heavy-fermion metals represent
a prototype case study for quantum criticality \cite{SiSteglich}.
Considerable theoretical work has emphasized the Kondo-breakdown
local quantum criticality \cite{Si-Nature,Coleman-JPCM}. Compared
with the spin-density-wave picture \cite{Hertz,Millis}, which is based
on the Landau notion of order-parameter fluctuations,
the Kondo breakdown
introduces
new low-energy
degrees of freedom.
The characteristic properties include, {\it e.g.},
a jump between large and
small Fermi surfaces \cite{Paschen,Shishido2005}.

Recently, experiments in YbRh$_{\rm 2}$Si$_{\rm 2}$ that is either
doped \cite{Friedemann09,Custers10} or pressurized \cite{Tokiwa09}
have revealed a rich phase diagram.
Under sufficient positive or negative (chemical)
pressure, the Kondo-breakdown point can be separated from the
AFM transition.
These results have been interpreted in terms of a global phase diagram,
which was put forward several years ago and more extensively discussed
recently \cite{Si_PhysicaB2006,YamamotoSi_JLTP2010,Coleman_JLTP2010}.

The global phase diagram emphasizes the interplay between two effects.
One is the Kondo screening and its breakdown, and the other the
fluctuations in the quantum magnetism of local moments alone.
The relevant zero-temperature phases can be either AFM
or paramagnetic, and can have ``large" or ``small" Fermi
surfaces \cite{Si_PhysicaB2006,YamamotoSi_JLTP2010,Coleman_JLTP2010}.
The large and small Fermi surfaces respectively correspond to the
cases with Kondo screening and destruction.
These results promise to take the study of heavy-fermion phase diagram
to an entirely new direction \cite{SiSteglich}.

A promising approach to the global phase diagram starts from
the AFM state, using a quantum non-linear sigma
model (QNL$\sigma$M) representation \cite{Yamamoto2007,YamamotoSi_JLTP2010}.
In dimensions higher than one, the Kondo coupling turns out to be exactly
marginal in the renormalization group (RG) sense,
and this shows a stable AFM phase
with Kondo destruction. Such a phase serves as
a basis
to describe different types of phase transitions
out of the AFM state \cite{Si_PhysicaB2006,YamamotoSi_JLTP2010}.
The low-energy physics in the ordered state involves only the smooth
space-time configurations of the sigma model field ${\bf n}$, for which
the spin Berry phase
vanishes. In order to access the Kondo-screened or otherwise paramagnetic
phases, topologically non-trivial configurations of the ${\bf n}$ field
will also be important; for such configurations, the spin Berry phase
is non-zero.

To gain insight into the effect of the Berry phase on the zero-temperature
phases of Kondo lattice systems, here we consider the case of
AFM spin-$1/2$ Kondo-Heisenberg model in one dimension.
We use the QNL$\sigma$M basis to study the effect of topological spin
excitations, with important guidance provided by the results derived from
bosonization method.
We show that,
when the conduction electron moves in the topologically nontrivial instanton
configurations of the ${\bf n}$ fields, a
Berry phase $\theta$ term
with $\theta_c=\pi$ arises.
The emergent Berry phase shifts the $\theta$ term of the spin chain
from $\pi$ to $0[mod~2\pi]$,
which in turn gives rise to a spin gap that is characteristic
\cite{AffleckWhite,Zachar,SiPivovarov}
of the Kondo-screened state of the one-dimensional Kondo-Heisenberg lattice.
Our results apply to both the insulating case at half-filling, where they are
consistent with the result of Tsvelik \cite{Tsvelik1994}, as well as the metallic
case away from half-filling.

{\it Kondo-Heisenberg model and QNL$\sigma$M mapping:} We
begin with the following one dimensional Hamiltonian
 \begin{eqnarray}
H=H_0+J_K\sum_{i}\mathbf{s}_{i}\cdot \boldsymbol \tau_i+J_H\sum_{i}\boldsymbol \tau_{i}\cdot \boldsymbol \tau_{i+1},
\end{eqnarray}
where $H_0=-t\sum_{i,\alpha}c_{i,\alpha}^{\dagger}c_{i+1,\alpha}+h.c.$, and the fermion spins and spin-half local moments are respectively described by $\mathbf{s}_i=\frac{1}{2}c_{i,\alpha}^{\dagger}\boldsymbol \sigma_{\alpha \beta}c_{i,\beta}$, and $\boldsymbol \tau_i$.The Kondo ($J_K$) and nearest neighbor ($J_H$) exchange  couplings are both antiferromagnetic,  and $t$ is the fermion hopping strength. We will work in the regime $J_K \ll J_H, t$, where we can use a continuum approximation for the spin chain. For the
latter, we first consider the semiclassical QNL$\sigma$M mapping\cite{Haldane}, followed by the bosonization method\cite{AffleckHaldane}.

In the semiclassical approximation we take $\boldsymbol \tau_i=(-1)^{i}S\mathbf{n}_i(1-\frac{a^2\mathbf{L}_{i}^{2}}{S^2})^{1/2}+a\mathbf{L}_{i}$,
where the unit vector field $\mathbf{n}_{i}$ is the staggered magnetization, and $\mathbf{L}_i$ is a canting field that satisfies the constraint $\mathbf{n}_i\cdot \mathbf{L}_i=0$. After integrating out $\mathbf{L}_i$ we obtain the effective action $S_{eff}=S[\mathbf{n}]+S_0+S_K$, where $S_0=\int d^2x \overline{\psi}_{s}[\gamma_{0}\partial_{0}+v\gamma_1\partial_1]\psi_{s}$, while $S[\mathbf{n}]$ and $S_K$ respectively describe the QNL$\sigma$M for the local moments and the Kondo interactions
The three terms
respectively describe the QNL$\sigma$M for the local moments, the free electrons, and the Kondo interactions:
\begin{eqnarray}
&&S[\mathbf{n}]=\frac{\rho_s}{2}\int d^2x [(\partial_1 \mathbf{n})^2+\frac{1}{c^2}(\partial_0 \mathbf{n})^2]+i \theta W[\mathbf{n}]\label{eq:2}\\
&&S_K=\int d^2x \bigg[\lambda_b e^{-i\pi r_j}\overline{\psi}e^{-i2k_Fr_j\gamma_5}\boldsymbol \sigma \psi \cdot \mathbf{n}\nonumber \\ && +i\lambda_f \overline{\psi}\gamma_0\boldsymbol \sigma \psi \cdot (\mathbf{n}\times\partial_0\mathbf{n})\bigg]+\ldots \label{eq:4}
\end{eqnarray}
Here the topological term $i \theta W[\mathbf{n}]$ corresponds to the Berry's phase due to the instanton configurations of $\mathbf{n}$, with $\theta=2\pi S$, and the Pontryagin index  $W[\mathbf{n}]=\frac{1}{4\pi}\int d^2x \ \mathbf{n}\cdot(\partial_x \mathbf{n} \times \partial_0 \mathbf{n})$ counts the winding number of the instantons. In the semiclassical approximation only the Berry's phase term retains the information about the quantized value of the spin. Based on this mapping, Haldane conjectured that half-integer spin chains characterized by $\theta=\pi$ are gapless, and in contrast the integer spin chains with $\theta=2\pi$ are gapped\cite{Haldane}.
In addition $\rho_s=J_HS^2a$, $c=2J_HSa$, respectively denote the spin stiffness and spin-wave velocity,
$v$ is the Fermi velocity, $k_F$ is the Fermi momentum, and the anticommuting gamma matrices are $\gamma_{0}=\eta_1$, $\gamma_1=\eta_2$ and $\gamma_5=i\gamma_0\gamma_1=-\eta_3$, with $\eta_i$'s being Pauli matrices. These Pauli matrices operate on the fermionic spinor
$\psi^{\dagger}=(R^{\dagger},L^{\dagger})$,
where $R$ and $L$ are the right and left moving fields,
and $\bar{\psi}=\psi^{\dagger}\gamma_0$.

The two terms in Eq.~\ref{eq:4} with coupling constants $\lambda_b$ and $\lambda_f$
respectively
 correspond to the backscattering and forward scattering interactions between the fermions and local moments;
 both
 are $\propto J_K$. The ellipsis indicates a four-fermion interaction term,
  obtained after integrating out $\mathbf{L}_{i}$, that is not important for our purpose.
  The backscattering term describes the coupling of the staggered magnetization densities of the electrons and spin chain,
  and the forward scattering term describes the coupling of the uniform magnetization densities. At half-filling,
  the product of the exponential phase factors is unity
  due to the commensurability of the conduction electrons and spin chain,
  and the backscattering term contributes as a relevant operator.
  Away from half-filling, the product is oscillatory in space, which
   makes the backscattering term irrelevant and the low energy physics is governed by the forward scattering term.

\paragraph{Bosonization results:}
   Before proceeding with the calculations within the QNL$\sigma$M approach, we will use bosonization method to gain insight into the Kondo singlet
formation \cite{Zachar,SiPivovarov}  and the emergent Berry phase. We show that the abelian-bosonization description of the Kondo-singlet state,
when transformed in terms of a non-abelian bosonization method, already suggests an emergent Berry phase.

In the bosonization approach the spin chain is first described in terms of fermions $\Psi$ with frozen charge fluctuations, and the Kondo interaction term is expressed as
\begin{eqnarray}
S_{K}=\int d^2x[\lambda_{b}\overline{\psi}e^{-i2k_Fr_j\gamma_5}\boldsymbol \sigma \psi \cdot \overline{\Psi}e^{-i\pi r_j\gamma_5}\boldsymbol \sigma \Psi \nonumber \\ +\lambda_{f}\overline{\psi}\gamma_0\boldsymbol \sigma \psi \cdot \overline{\Psi}\gamma_0\boldsymbol \sigma \Psi]. \label{eq:5}
\end{eqnarray}
Within the Abelian bosonization, the collective charge and spin fluctuations of the electrons are respectively described by the bosonic fields $\varphi_c$, $\varphi_{s}$ and their corresponding dual fields $\theta_c$, $\theta_s$. The spin fluctuations of the local moments are described by the bosonic field $\varphi_{\tau}$ and its dual $\theta_{\tau}$. The kinetic energy of the fermions are described in terms of Gaussian actions involving $\varphi_c$, $\varphi_s$, and $\varphi_{\tau}$,
and $S_K$ becomes
\begin{eqnarray}
S_K&\propto & \int d^2x \bigg [\lambda_b \cos((2k_F+\pi) r_j+\sqrt{2\pi}\varphi_c)(\cos \sqrt{2\pi}\varphi_{-} \nonumber \\&&-\cos \sqrt{2\pi}\varphi_{+} +2\cos \sqrt{2\pi}\theta_{-})+\lambda_{f}\partial_x\varphi_s\partial_x\varphi_{\tau} \nonumber \\&&+\lambda_{f}\cos \sqrt{2\pi}\theta_{-}(\cos \sqrt{2\pi}\varphi_{-}+\cos \sqrt{2\pi}\varphi_{+})\bigg ],
\end{eqnarray}
where $\varphi_{\pm}=\varphi_{s}\pm \varphi_{\tau}$ and $\theta_{-}=\theta_{s}-\theta_{\tau}$.

Away from the half filling,
the low
energy physics is controlled by the forward scattering term that is marginally
relevant  \cite{SiPivovarov} .
The $\cos \sqrt{2\pi}\theta_{-}\cos \sqrt{2\pi}\varphi_{-}$ coupling is irrelevant
and can be ignored.
Since the forward scattering operators do not couple the charge and spin sectors,
the charge field remains a free field and leads to the metallic behavior in the
charge sector. However the spin fields satisfy one of the
following locking conditions:
\begin{eqnarray}
\sqrt{2\pi}\varphi_+=2n_1\pi, \ \sqrt{2\pi}\theta_-=(2n_2+1)\pi \label{eq:7}
 \\
\sqrt{2\pi}\varphi_+=(2n_1+1)\pi, \ \sqrt{2\pi}\theta_-=2n_2\pi ,
\label{eq:8}
\end{eqnarray}
which signals the Kondo singlet formation and an emergent
spin gap \cite{Zachar, SiPivovarov, Berg}. As a result of the Kondo singlet
formation there is a gapless charge density wave mode
at wavevector $2k_{F}^{\ast}=2k_F+\pi$, described by $\langle \mathbf{N}^{\tau} \cdot \mathbf{N}^{s} \rangle \propto \cos\left((2k_F+\pi)r_j+\sqrt{2\pi}\varphi_c\right)$.

The above can be compared with the insulating system at half filling,
where the backscattering and forward scattering operators are respectively
relevant and marginally relevant operators.
Consequently\cite{AffleckWhite,Zachar, SiPivovarov}, the low energy
physics is governed by the backscattering term, a potential energy
of the form $\mathbf{N}^{\tau} \cdot \mathbf{N}^{s}$, where
$\mathbf{N}_{s,\tau}$ are the staggered magnetization densities.
This form implies that
the spin fields will lock into a configuration such that
$\mathbf{N}^{\tau} \cdot \mathbf{N}^{s}=-1$. Combining the energy minimum
criterion with the fact that $\cos \sqrt{2\pi}\varphi_-$ and
$\cos \sqrt{2\pi}\theta_-$ can not have simultaneous vacuum expectation values,
we find two possibilities for the charge and spin fields,
either with Eq.~(\ref{eq:7}) and $\sqrt{2\pi}\varphi_c=2n_3\pi$,
or with Eq.~(\ref{eq:8}) and
$\sqrt{2\pi}\varphi_c=(2n_3+1)\pi$.
The nonzero
$\langle\cos \sqrt{2\pi}\varphi_{c}\rangle$ causes a charge gap,
leading to a charge insulator behavior.
In the spin sector, the above reveals
an important insight, which appears not to have been appreciated before:
the locking conditions for the spin bosons
in the half-filled insulating case and
away-from-half-filled metallic cases are identical.
This insight will be important in guiding our subsequent analysis of
the Berry phase effect in the
QNL$\sigma$M representation.

To anticipate the QNL$\sigma$M analysis, we now
demonstrate the relation between the Kondo singlet formation and  topological Berry phase
using the non-Abelian bosonization
method\cite{AffleckHaldane}. The spin sector for the electrons
and local moments are described by the $SU(2)$ matrix fields $\mathcal{U}_{s,\tau}$ and the
corresponding $SU_1(2)$ Wess-Zumino-Witten (WZW) actions
\begin{eqnarray}
S_{s,\tau}=\frac{1}{16\pi}\int d^2x Tr[\partial_{\mu}\mathcal{U}_{s,\tau}^{\dagger}\partial_{\mu}\mathcal{U}_{s,\tau}]+\Gamma_{WZ}[\mathcal{U}_{s,\tau}]\nonumber \\
\Gamma_{WZ}[\mathcal{U}_{s,\tau}]= \frac{-i}{24\pi}\int_{0}^{\infty} d\xi \int d^2x \epsilon_{\alpha \beta \gamma} Tr[\Omega_{s,\tau}^{\alpha}\Omega_{s,\tau}^{\beta}\Omega_{s,\tau}^{\gamma}],\nonumber \\
\end{eqnarray}
where $\Gamma_{WZ}$ is the topological Wess-Zumino (WZ) term and $\Omega_{s,\tau}^{\alpha}=\mathcal{U}_{s,\tau}^{\dagger}\partial_{\alpha}\mathcal{U}_{s,\tau}$. For the electrons and local moments the space-time coordinates are respectively described by $(v_{s,\tau}x_0,x_1)$. The topological WZ term is crucial to maintaining the gapless behavior of the $\mathcal{U}_{s,\tau}$ fields. The matrix fields can be decomposed as $\mathcal{U}_{s, \tau}=u_{0,s, \tau}+i\mathbf{u}_{s, \tau}\cdot \boldsymbol \sigma$, with $u_{0,s, \tau}^{2}+\mathbf{u}_{s, \tau}^{2}=1$, where $u_{0,s, \tau}$, and $\mathbf{u}_{s, \tau}$ respectively describe singlet spin-Peierls and staggered magnetization correlations. The relationship among the non-Abelian and Abelian bosonization fields are described by
\begin{eqnarray}
u_{0,s}\pm i u_{3,s}=e^{\pm i\sqrt{2\pi}\phi_s},  \ u_{1,s}\pm iu_{2,s} =
\pm
ie^{\mp i\sqrt{2\pi} \theta_s}\nonumber \\
\end{eqnarray}
The locking conditions of the Abelian fields,
Eqs.~(\ref{eq:7},\ref{eq:8}),
translate into $\mathcal{U}_s=\pm \mathcal{U}_{\tau}^{\dagger}$. Using the property $\Gamma_{WZ}[\mathcal{U}^{\dagger}]=-\Gamma_{WZ}[U]$, we find that the Kondo-singlet formation is accompanied by the cancellation between the WZ terms of the spin chain and the electrons. This leaves an effective matrix sigma model without the topological term, which is known to be gapped. If the Peierls type singlet correlation $u_0$ is suppressed, the WZW action reduces to a QNL$\sigma$M, and the WZ term transforms into a topological $\theta=\pi$ Berry phase for the QNL$\sigma$M. Therefore we anticipate that the Kondo singlet formation within a sigma model approach will be associated with an emergent $\theta=\pi$ Berry phase from the electronic part of the action.

\paragraph{QNL$\sigma$M at half-filling :} After gaining insight into the Kondo-singlet formation via bosnization analysis, we turn to the QNL$\sigma$M approach. At half-filling the problem can be solved in an elegant manner due to Tsvelik\cite{Tsvelik1994}. Introducing the non-Abelian bosonization field $\mathcal{U}_s$ for the conduction electrons, the relevant back scattering term can be expressed as $4\lambda_b\mathbf{u}_s\cdot\mathbf{n}\cos \sqrt{2\pi}\varphi_c$. The energy minimization is achieved for $\mathbf{u}_s=\mathbf{n}$, $\sqrt{2\pi}\varphi_c=(2n+1)\pi$ or $\mathbf{u}_s=-\mathbf{n}$, $\sqrt{2\pi}\varphi_c=2n\pi$. The conditions $\mathbf{u}_s=\pm\mathbf{n}$ imply $u_{0,s}=0$, and the WZ term becomes $\pm i\pi W[\mathbf{n}]$ which cancels the $\theta=\pi$ Berry phase term of the spin chain. Consequently we obtain charge and spin gaps.
However this approach does not account for the forward scattering terms and can not be applied to the metallic case away from half filling, where
a new treatment is required.

\paragraph{Berry's phase from non-Abelian chiral anomaly:}
We now analyze the Kondo effect and emergent Berry phase in the
QNL$\sigma$M representation at arbitrary filling.
Based on the bosonization results,
we
see that Kondo singlet formation is accompanied by the $2k_{F}^{\ast}$ charge
density wave oscillation described by $\langle \mathbf{N}_s\cdot\mathbf{N}_{\tau}\rangle$. Therefore in the sigma model approach we need to find an appropriate fermionic basis such that the component of $\mathbf{N}_s$ parallel to $\mathbf{n}$ has $2k_F^{\ast}$ charge oscillation (at half-filling due to commensurability $2k_{F}^{\ast}$ charge mode remains gapped).

Recognizing that $\mathbf{N}_s$ contains the combination of left and right moving fields, we anticipate that a spin dependent chiral transformation will be needed to describe the appropriate fermionic basis. In the following section we demonstrate that both at and away from half-filling the emergent Berry's phase can be calculated by using a non-Abelian chiral rotation and the associated chiral anomaly \cite{Fujikawa, GamboaSaravi, Tanaka}.
In Ref.~\onlinecite{Tanaka} the non-Abelian chiral rotation technique has been applied to calculate the Berry phase at half-filling in the absence of the forward scattering term. However the relation between the emergent Berry phase and Kondo singlet formation has not been addressed. We consider this relationship and, in addition, study the forward scattering term to address the metallic case away from half-filling.

We perform a spin-dependent chiral rotation $\psi \to \exp(i\phi \mathbf{n}\cdot \boldsymbol \sigma \gamma_5)\chi$. The staggered magnetization transforms into
\begin{eqnarray}
\overline{\psi}e^{-i2k_Fr_j\gamma_5}\frac{\boldsymbol \sigma}{2} \psi&=& \frac{1}{2}\bar{\chi}[\boldsymbol \sigma-\mathbf{n}(1-\cos 2\phi)\mathbf{n}\cdot \boldsymbol \sigma \nonumber \\ & &+i\gamma_5\sin 2\phi \mathbf{n}]e^{-i2k_Fr_j\gamma_5}\chi
\end{eqnarray}
After taking a dot product with $\mathbf{n}$, we find that only for $\phi=\pm \pi/4$, $\mathbf{N}_s\cdot \mathbf{n}$ demonstrates pure charge density oscillations with $2k_F^{\ast}$ wavevector. Therefore $\phi=\pi/4$ is the required chiral rotation angle, which removes the spin dependence of the back scattering term and converts it into $i\lambda_b\exp(-i\pi r_j-2ik_Fr_j\gamma_5)\bar{\chi}\gamma_5\chi$. Therefore at half-filling, the back scattering term becomes $i\lambda_b\bar{\chi}\gamma_5\chi$, and causes a charge gap. One can also perform a successive $U(1)$ chiral rotation $\exp(-i\frac{\pi}{4}\gamma_5)\chi$, to transform $i\bar{\chi}\gamma_5\chi$ into a ordinary mass term $\lambda_b \bar{\chi}\chi$. However this is not necessary for the physics in the spin-sector.

Since the functional measure is not invariant under chiral rotation,
we need to find the Jacobian of the transformation which leads
to the chiral anomaly terms. After an explicit calculation detailed
in the supplementary materials \cite{supplementary}, we find the Jacobian
\begin{eqnarray}
J\left(\phi=\frac{\pi}{4}\right)&=&\exp\bigg[-i\pi W[\mathbf{n}]+\int d^2x \bigg\{\frac{v}{4\pi}(\partial_1 \mathbf{n})^2\nonumber \\
&&+\frac{(1-2\lambda_f)^2}{4\pi v}(\partial_0 \mathbf{n})^2\bigg\}\bigg]
\end{eqnarray}
The Jacobian consists an emergent $\theta=\pi$ Berry phase term, and two additional terms which renormalize the spin stiffness and spin wave velocity of the QNL$\sigma$M. The emergent Berry phase cancels the existing $\pi$ Berry phase of the spin chain, and renders the sigma model field gapped. As a result of the $\pi/4$ chiral rotation, the spin sector of the $\chi$ fermions also becomes gapped. This can be demonstrated by considering the bosonization of the $\chi$ fermions. In the non-abelian bosonization formulation, the spin sector of the $\chi$ fermion does not contain the topological WZW term, and the matrix sigma model becomes gapped. The fermionic action transforms into
\begin{eqnarray}
S_{f}=\int d^2x \bar{\chi}[\gamma_{\mu}\partial_{\mu}+\frac{i}{2}\gamma_{\mu}\gamma_5\boldsymbol \sigma \cdot \partial_{\mu}\mathbf{n}+\frac{i}{2}\gamma_{\mu}\boldsymbol \sigma \cdot (\mathbf{n}\times \partial_{\mu}\mathbf{n})\nonumber \\
-i\lambda_f\gamma_{0}\gamma_5\boldsymbol \sigma \cdot \partial_{0}\mathbf{n}+i\lambda_be^{-i(\pi +2k_F)r_j\gamma_5}\gamma_5]\chi +\ldots \nonumber \\
\end{eqnarray}
Since the $\mathbf{n}$ field, and the spin sector of $\chi$ field are gapped, the interaction between these fields describe the innocuous fluctuations about Kondo singlet phase.

To summarize, the spin dependent chiral rotation by angle $\pi/4$
incorporates the Kondo singlet formation.
In the metallic case away from half-filling,
a $\theta=\pi$ Berry phase emerges
as a consequence of chiral anomaly.
The effects of the Berry phase and instantons in the spin sector turn out to be the same as those at half-filling.
The difference between the two cases exists only in the charge sector, and the term causing the
gap at half-filling no longer operates away from half-filling.

\paragraph{Competition with spin-Peierls correlations:}
The emergent theta term highlights the role of instanton configurations of the $\mathbf{n}$ field.
We now discuss its connection with the spin-Peierls order parameter.
In the semiclassical language spin Peierls order
parameter $(-1)^i\langle\mathbf{S}_{i}\cdot \mathbf{S}_{i+1}\rangle$ corresponds to the instanton density $a^2\mathbf{n}\cdot(\partial_x \mathbf{n} \times \partial_0 \mathbf{n})$. Therefore the instantons of the sigma model are manifestations of the competition between
the spin Peierls and N\'{e}el order. For the spin one-half case we consider, $\theta=\pi$ and
 the gaplessness of the spin chain implies the same power law correlation of the Peierls and the N\'{e}el order parameters. 
The Kondo singlet formation is detrimental to both singlet Peierls and triplet N\'{e}el correlations,
as the system moves away from the gapless point; this competition between two types of singlet correlations is encoded in the emergent $\pi$ Berry phase term.

This conclusion demonstrates the effect of the emergent Berry phase beyond the description of
how Kondo-singlet paramagnetic phase transitions out of a Kondo-breakdown spin-liquid reference point.
The Berry phase also characterizes the competition between the Kondo paramagnet and Kondo-breakdown spin-Peierls phase.
While the Fermi momenta of the Kondo-singlet paramagnet are large, those of the spin-Peierls state are small.
These paramagnetic phases and their transitions resemble the paramagnetic portion of the global phase
diagram that has been proposed for heavy-fermion metals.

In higher dimensions there is a stable N\'{e}el ordered AF state, and the instantons are suppressed by finite spin stiffness in the magnetically ordered phase. However in the quantum disordered region the consideration of the instantons becomes relevant, and the associated Berry phase is critical in determining the nature of the emergent valence bond solid phase\cite{Duncan,Read}. Therefore it is conceivable that Berry phase effects related to what we have considered here will be important in dimensions higher than one.

We acknowledge the support of NSF Grant No. DMR-1006985, the Robert A. Welch Foundation Grant No. C-1411, and the W.\ M.\ Keck Foundation.

\pagebreak

\onecolumngrid

\section*{Supplementary Material for EPAPS}

\section{I. Details of the chiral anomaly calculation described in the main text}

Following Ref.~\onlinecite{Fujikawa, GamboaSaravi, Tanaka} we build up the finite chiral rotation by angle $\pi/4$, by applying successive chiral rotations at infinitesimal increment: $\psi \to \exp(i\frac{\pi}{4}\alpha  \mathbf{n}\cdot \boldsymbol \sigma \gamma_5)\chi$, where $\alpha $ is the increment parameter that ranges between zero and unity. Under such transformation, the forward scattering term and the kinetic energy parts respectively transform into
\begin{eqnarray}
i\lambda_f\bar{\psi}\gamma_0 \boldsymbol \sigma \cdot (\mathbf{n}\times \partial_0\mathbf{n})\psi \to i\lambda_f \bar{\chi}\left[ \gamma_0\boldsymbol \sigma \cdot (\mathbf{n}\times \partial_0\mathbf{n}) \cos \frac{\pi \alpha}{2} -\gamma_0 \gamma_5 \partial_0\mathbf{n}\cdot \sigma \sin \frac{\pi \alpha}{2}\right]\chi\\
\bar{\psi}\gamma_{\mu}\partial_{\mu}\psi  \to  \bar{\chi}\left[\gamma_{\mu}\partial_{\mu}+\frac{i}{2}\sin \frac{\pi \alpha}{2}\gamma_{\mu}\gamma_5 \partial_{\mu}\mathbf{n}\cdot \sigma +\frac{i}{2}\left(1-\cos \frac{\pi \alpha}{2}\right)(\mathbf{n}\times \partial_{\mu}\mathbf{n})\cdot \sigma\gamma_{\mu}\right]\chi
\end{eqnarray}
where we have set the Fermi velocity $v=1$, and at the end of the calculations we will restore $v$ by dimensional analysis.
The Jacobian is evaluated as
\begin{eqnarray}
J\left(\phi=\frac{\pi}{4}\right)=\exp\left \{\frac{i}{8\pi}\int_{0}^{1} d\alpha \int d^2x Tr\left[\pi \mathbf{n} \cdot \boldsymbol \sigma \gamma_5 D_{\alpha}^{2}\right]\right \}
\end{eqnarray}
where the Dirac operator $D_{\alpha}$ is found by combining all the transformed fermion bilinears. The explicit expression for $D_{\alpha}$ is given by
\begin{eqnarray}
D_{\alpha}&=&\gamma_{\mu}\left[\partial_{\mu}+\frac{i}{2}\sin \frac{\pi \alpha}{2}\gamma_5 \partial_{\mu}\mathbf{n}\cdot \sigma +\frac{i}{2}\left(1-\cos \frac{\pi \alpha}{2}\right)(\mathbf{n}\times \partial_{\mu}\mathbf{n})\cdot \sigma\right]+\lambda_be^{-i(\pi+2k_F)r_j\gamma_5}\bigg[\cos \frac{\pi \alpha}{2}\mathbf{n}\cdot \boldsymbol \sigma +i\sin \frac{\pi \alpha}{2}\gamma_5\bigg]\nonumber \\ &&+i\lambda_f \gamma_0 \left[ \boldsymbol \sigma \cdot (\mathbf{n}\times \partial_0\mathbf{n}) \cos \frac{\pi \alpha}{2}- \gamma_5 \partial_0\mathbf{n}\cdot \sigma \sin \frac{\pi \alpha}{2}\right]
\end{eqnarray}
To extract the non-vanishing contributions to the Jacobian, we need to look for only those terms in $D_{\alpha}^{2}$, which contain the matrices $\gamma_5$ and $\boldsymbol \sigma$, and only the terms proportional to $(\partial_0 \times \partial_1 \mathbf{n})\cdot \sigma$ lead to the topological theta term. From the $D_{\alpha}^2$ the following four terms are found to contribute to the topological Berry phase
\begin{eqnarray}
&&f_1=-\frac{i}{8\pi}\times \frac{\pi}{2}\int_{0}^{1}d\alpha \int d^2x Tr[\mathbf{n}\cdot \boldsymbol \sigma \gamma_5\gamma_5(\partial_0\mathbf{n}\times \partial_1\mathbf{n})\cdot \boldsymbol \sigma ]\left(1-\cos \frac{\pi \alpha}{2}\right)=-i\pi W[\mathbf{n}]\left(1-\frac{2}{\pi}\right) \\
&&f_2=-\frac{i}{8\pi}\times \frac{\pi}{2}\int_{0}^{1}d\alpha \int d^2x Tr[\mathbf{n}\cdot \boldsymbol \sigma \gamma_5\gamma_5(\partial_0\mathbf{n}\times \partial_1\mathbf{n})\cdot \boldsymbol \sigma ]\left(1-\cos \frac{\pi \alpha}{2}+2\lambda_f\cos \frac{\pi \alpha}{2}\right)= -i\pi W[\mathbf{n}]\left(1-\frac{2}{\pi}+\frac{4\lambda_f}{\pi}\right) \nonumber \\ \\
&&f_3=-\frac{i}{8\pi}\times \frac{\pi}{4}\times (1-2\lambda_f)\int_{0}^{1}d\alpha \int d^2x Tr[\mathbf{n}\cdot \boldsymbol \sigma \gamma_5\gamma_5(\partial_0\mathbf{n}\times \partial_1\mathbf{n})\cdot \boldsymbol \sigma ]\left(1-\cos\pi \alpha\right)=-i\frac{\pi}{2} W[\mathbf{n}]\left(1-2\lambda_f\right) \\
&&f_4=\frac{i}{8\pi}\times \frac{\pi}{2}\int_{0}^{1}d\alpha \int d^2x Tr[\mathbf{n}\cdot \boldsymbol \sigma \gamma_5\gamma_5(\partial_0\mathbf{n}\times \partial_1\mathbf{n})\cdot \boldsymbol \sigma ]\left(1-\cos \frac{\pi \alpha}{2}\right)\left(1-\cos \frac{\pi \alpha}{2}+2\lambda_f\cos \frac{\pi \alpha}{2}\right)\nonumber \\
&&=i\pi W[\mathbf{n}]\left(\frac{3}{2}-\frac{4}{\pi}+\frac{4\lambda_f}{\pi}-\lambda_f\right)
\end{eqnarray}
After combining all four contributions we obtain the net Berry phase $f_1+f_2+f_3+f_4=-i\pi W[\mathbf{n}]$. For an arbitrary chiral rotation by angle $\phi$, the net Berry phase due to chiral anomaly can be calculated in a similar manner and the final answer turns out to be $-iW[\mathbf{n}](4\phi -(1-2\lambda_f)\sin 4\phi)$. Only for two particular values $\phi=\pi/4, \ \pi/2$, the Berry phase turns out to be independent of $\lambda_f$. These two angles respectively give $-\pi, \ -2\pi$ Berry phase. As we have explained in the main text only $\phi=\pi/4$ rotation is consistent with the appearance of pure $2k_{F}^{\ast}$ charge density oscillation of $\langle N_s \cdot N_{\tau}\rangle$.

The Jacobian also has following four nontopological contributions
\begin{eqnarray}
&&g_1=\frac{i}{8\pi}\times \frac{i\pi}{2v}\times (1-2\lambda_f)\int_{0}^{1}d\alpha \int d^2x Tr[\mathbf{n}\cdot \boldsymbol \sigma \gamma_5\gamma_5\partial_{0}^{2}\mathbf{n}\cdot \boldsymbol \sigma ]\sin \frac{\pi \alpha}{2}=\frac{(1-2\lambda_f)}{2\pi v}(\partial_0\mathbf{n})^2 \\
&&g_2=\frac{i}{8\pi}\times \frac{iv\pi}{2}\int_{0}^{1}d\alpha \int d^2x Tr[\mathbf{n}\cdot \boldsymbol \sigma \gamma_5\gamma_5\partial_{1}^{2}\mathbf{n}\cdot \boldsymbol \sigma ]\sin \frac{\pi \alpha}{2}=\frac{v}{2\pi}(\partial_1\mathbf{n})^2 \\
&&g_3=\frac{i}{8\pi}\times \frac{i\pi}{2v}\times (1-2\lambda_f)\int_{0}^{1}d\alpha \int d^2x Tr[\mathbf{n}\cdot \boldsymbol \sigma \gamma_5\gamma_5\mathbf{n}\cdot \boldsymbol \sigma ](\partial_0\mathbf{n})^2\sin \frac{\pi \alpha}{2}\left(1-\cos \frac{\pi \alpha}{2}+2\lambda_f\cos \frac{\pi \alpha}{2}\right)\nonumber \\&&=-\frac{(1-4\lambda_{2}^{2})}{4\pi v}(\partial_0\mathbf{n})^2 \\
&&g_4=\frac{i}{8\pi}\times \frac{iv\pi}{2}\int_{0}^{1}d\alpha \int d^2xTr[\mathbf{n}\cdot \boldsymbol \sigma \gamma_5\gamma_5\mathbf{n}\cdot \boldsymbol \sigma ](\partial_1\mathbf{n})^2 \sin \frac{\pi \alpha}{2}\left(1-\cos \frac{\pi \alpha}{2}\right)=-\frac{v}{4\pi}(\partial_1\mathbf{n})^2 \end{eqnarray}
After combining all four nontopological contributions we obtain
\begin{equation}
g_1+g_2+g_3+g_4=\frac{v}{4\pi}(\partial_1\mathbf{n})^2 +\frac{(1-2\lambda_{2})^2}{4\pi v}(\partial_0\mathbf{n})^2,
\end{equation}
which leads to the renormalization of the spin stiffness constant and the spin wave velocity of the QNL$\sigma$M.

\section{II. Anomaly calculation in a diagonally $SU(2)$ rotated basis} In the literature the spin-fermion interaction is often treated using a popular $SU(2)$ rotated basis in which the spin quantization axis of the fermions is locally rotated along the QNL$\sigma$M field $\mathbf{n}$. This rotation is accomplished by the following unitary transformation $\psi \to U \psi$, where $U^{\dagger}\mathbf{n}\cdot \sigma U=\sigma_3$. In this section we begin with this rotated basis, and
demonstrate that a subsequent chiral rotation yields the same $\theta=\pi$
Berry phase
as a consequence of the chiral anomaly.
The calculation in such a rotated basis has been performed
in Ref.~\onlinecite{Tanaka}, but
the authors arrived at an incorrect conclusion that the non-Abelian chiral
anomaly vanishes in this basis. In the following we demonstrate the physically sensible result that
the chiral anomaly is insensitive to arbitrary diagonal or non-chiral
$SU(2)$ rotations.
Under the $SU(2)$ rotation described above, the backscattering term
transforms into
\begin{equation}
\lambda_b\int d^2x \overline{\psi}e^{-i(\pi+2k_F)r_j\gamma_5}\boldsymbol \sigma  \psi \cdot \mathbf{n}\to \lambda_b\int d^2x \overline{\psi}e^{-i(\pi+2k_F)r_j\gamma_5}\sigma_3\psi
\end{equation}
and the fermion kinetic energy transforms into
\begin{equation}
\int d^2x \bar{\psi}\gamma_{\mu}\partial_{\mu}\psi \to \int \bar{\psi}\gamma_{\mu}\left(\partial_{\mu}+\frac{i}{2}\mathcal{A}_{\mu}^{a}\sigma_a\right)\psi
\end{equation}
where we have introduced the $SU(2)$ vector potentials $\mathcal{A}_{\mu}^{a}$, using the relation $U^{\dagger}\partial_{\mu}U=-\partial_{\mu}U^{\dagger}U=i/2\mathcal{A}_{\mu}^{a}\sigma_a$. The transformed forward scattering term is calculated as
\begin{eqnarray}
i\lambda_f\int d^2 x \bar{\psi}\gamma_0(\mathbf{n} \times \partial_0 \mathbf{n})\cdot \boldsymbol \sigma \psi=\lambda_f\int d^2 x \bar{\psi}\gamma_0(\mathbf{n}\cdot \boldsymbol \sigma \partial_0 \mathbf{n}\cdot \boldsymbol \sigma)\psi
\to \lambda_f \int d^2x \bar{\psi}\gamma_0\sigma_3(U^{\dagger}\partial_0U\sigma_3+\sigma_3\partial_0U^{\dagger}U)\psi \nonumber \\
=-i\lambda_f \int d^2x \bar{\psi} \gamma_0[\mathcal{A}_{0}^{1}\sigma_1+A_{0}^{2}\sigma_2]\psi
\end{eqnarray}
Therefore the transformed quadratic part of the fermionic action can be expressed as
\begin{equation}
S_{F,quad}=\int d^2x \bar{\psi}\left[\gamma_{\mu}\left(\partial_{\mu}+\frac{i}{2}\mathcal{A}_{\mu}^{a}\sigma_a\right)+\lambda_be^{-i(\pi+2k_F)r_j\gamma_5}\sigma_3-i\lambda_f\gamma_0\left(\mathcal{A}_{0}^{1}\sigma_1+\mathcal{A}_{0}^{2}\sigma_2\right)\right]\psi
\end{equation}
Now we need to perform the finite chiral rotation $\psi \to e^{i\pi\sigma_3\gamma_5/4}\chi$, to eliminate the spin dependence of the backscattering term. After the chiral rotation by the angle $\pi \alpha/4$, we obtain the following Dirac operator
\begin{eqnarray}
D_{\alpha}&=&\gamma_{\mu}\left[\partial_{\mu}+\frac{i}{2}\mathcal{A}_{\mu}^{3}\sigma_3+\frac{i}{2}\left(\mathcal{A}_{\mu}^{1}\sigma_1+\mathcal{A}_{\mu}^{2}\sigma_2\right)\cos \frac{\pi \alpha}{2}+\gamma_5\frac{i}{2}\left(\mathcal{A}_{\mu}^{1}\sigma_2-\mathcal{A}_{\mu}^{2}\sigma_1\right)\sin \frac{\pi \alpha}{2}\right]-i\gamma_0\bigg[\left(\mathcal{A}_{0}^{1}\sigma_1+\mathcal{A}_{0}^{2}\sigma_2\right)\cos \frac{\pi \alpha}{2}\nonumber \\
&&+\gamma_5\left(\mathcal{A}_{0}^{1}\sigma_2-\mathcal{A}_{0}^{2}\sigma_1\right)\sin \frac{\pi \alpha}{2}\bigg]+i\lambda_be^{-i(\pi+2k_F)r_j\gamma_5}
\end{eqnarray}
At this stage it becomes convenient to use the $CP^{1}$ representation of the sigma model field. By introducing a complex two component bosonic spinor $z^{T}=(z_{\uparrow},z_{\downarrow})$, with the constraint $z^{\dagger}z=1$, we can express the QN$\sigma$M field as $\mathbf{n}=z^{\dagger}\sigma z$. In the $CP^1$ representation the matrix $U$ becomes
\begin{equation}
U=\left[ {\begin{array}{cc}
 z_{\uparrow} & -z^{\ast}_{\downarrow}  \\
 z_{\downarrow} & \ \ \ z^{\ast}_{\uparrow}  \\
 \end{array} } \right]
\end{equation}
and the sigma model action is transformed into the following $CP^{1}$ action
\begin{equation}
S[z,\mathcal{A}_{\mu}^{3}]=\int d^2x\left[ \frac{2}{g}|(\partial_{\mu}-i\mathcal{A}_{\mu}^{3})z|^2+i\lambda(|z|^2-1)+i\frac{\theta}{4\pi}\epsilon_{\mu \nu}\partial_{\mu}\mathcal{A}_{\nu}^{3}\right].
\end{equation}
where $\theta=2\pi S$, and $\lambda$ is a Lagrange multiplier, and the dimensionless coupling constant $g=c/\rho_s$. In the $CP^1$ representation the topological theta term is expressed in terms of the electric field of the $U(1)$ gauge field $\mathcal{A}_{\mu}^{3}$. From the identity $U^{\dagger}U=1$, it follows that $\epsilon_{\mu \nu}\partial_{\mu}\mathcal{A}_{\nu}^{3}=(\mathcal{A}_{0}^{1}\mathcal{A}_{1}^{2}-\mathcal{A}_{0}^{2}\mathcal{A}_{1}^{1})$. This identity turns out to be important for the evaluation of the topological contributions to the Jacobian.
The Jacobian is evaluated as
\begin{eqnarray}
J(\alpha=1)=\exp\left[\frac{i}{8\pi}\int_{0}^{1} d\alpha \int d^2x Tr[\pi \sigma_3 \gamma_5 D_{\alpha}^{2}]\right]
\end{eqnarray}
To find the topological contributions we only pick the terms which are proportional to $\epsilon_{\mu \nu}\partial_{\mu}\mathcal{A}_{\nu}^{3}\sigma_3\gamma_5$ and $(\mathcal{A}_{0}^{1}\mathcal{A}_{1}^{2}-\mathcal{A}_{0}^{2}\mathcal{A}_{1}^{1})\sigma_3\gamma_5$. From $D_{\alpha}^{2}$, the following two terms are found to contribute to the topological Berry phase
\begin{eqnarray}
&&F_1=-\frac{i}{8\pi}\times \frac{\pi}{2}\int_{0}^{1} d\alpha \int d^2x Tr[\sigma_3 \sigma_3 \gamma_5 \gamma_5]\epsilon_{\mu \nu}\partial_{\mu}\mathcal{A}_{\nu}^{3}=-\frac{i \pi}{4\pi} \int d^2x \epsilon_{\mu \nu}\partial_{\mu}\mathcal{A}_{\nu}^{3} \\
&&F_2=\frac{i}{8\pi} \times \frac{\pi}{2}\times (1-2\lambda_f) \int_{0}^{1} d\alpha \int d^2x Tr[\sigma_3 \sigma_3 \gamma_5 \gamma_5](\mathcal{A}_{0}^{1}\mathcal{A}_{1}^{2}-\mathcal{A}_{0}^{2}\mathcal{A}_{1}^{1})\cos \pi \alpha=0 \label{eq:F2}
\end{eqnarray}
Therefore the net Berry phase term is $F_1+F_2=-\frac{i \pi}{4\pi} \int d^2x \epsilon_{\mu \nu}\partial_{\mu}\mathcal{A}_{\nu}^{3}=-i\pi W[\mathbf{n}]$. In Ref.~\onlinecite{Tanaka} the forward scattering term $\propto \lambda_f$ was not considered and $\cos \pi \alpha$ in Eq.~\ref{eq:F2} was replaced by unity. Consequently the Berry phase contribution from the chiral anomaly was found to vanish. For a general chiral rotation $\phi$, we find the net Berry phase  $-\frac{i}{4\pi}4\phi \int d^2x \epsilon_{\mu \nu}\partial_{\mu}\mathcal{A}_{\nu}^{3}+\frac{i}{4\pi}(1-2\lambda_f)\sin 4\phi \int d^2x (\mathcal{A}_{0}^{1}\mathcal{A}_{1}^{2}-\mathcal{A}_{0}^{2}\mathcal{A}_{1}^{1})=-iW[\mathbf{n}](4\phi -(1-2\lambda_f)\sin 4\phi)$, which agrees with the result of the previous section.

\end{document}